\documentclass[aps,prl,twocolumn,showpacs,superscriptaddress,groupedaddress]{revtex4}
\usepackage{amsmath,amssymb,color,graphicx,hyperref,amsthm,bm,epstopdf,mathrsfs}
\usepackage{placeins}

\renewcommand{\ol}{\overline} 
 
\renewcommand{\Re}{\mathrm{Re}} 
\renewcommand{\Im}{\mathrm{Im}}
\newcommand{\BB}{\mathcal{B}} 
\newcommand{\CC}{\mathcal{C}} 
\renewcommand{\AA}{\mathcal{A}} 

\newcommand{\nv}{{\bf n}}

\newcommand{\R}{{\mathbb{R}}}

\newcommand{\beq}{\begin{equation}}
\newcommand{\eeq}{\end{equation}}
\newcommand{\bea}{\begin{eqnarray}}
\newcommand{\eea}{\end{eqnarray}}
\newcommand{\ben}{\begin{eqnarray*}}
\newcommand{\een}{\end{eqnarray*}}
\newcommand{\bem}{\begin{enumerate}}
\newcommand{\eem}{\end{enumerate}}
\newcommand{\ra}{\rightarrow}

\newcommand{\cd}{\partial}

\newcommand{\ignore}[1]{}

\renewcommand{\star}{*}

\newcommand{\vphi}{\varphi}

\newcommand{\nvec}{\mbox{\boldmath{$n$}}}
\renewcommand{\nv}{\mbox{\boldmath{$n$}}}

\newcommand{\eps}{\varepsilon}

\definecolor{TW-color}{RGB}{100,0,100}
\definecolor{Error-color}{RGB}{250,0,0}

\begin{document}

\title{Chiral $p$-wave superconductors have complex coherence and magnetic field penetration lengths}
\author{Martin Speight}
\affiliation{School of Mathematics, University of Leeds, Leeds LS2 9JT, United Kingdom}
\author{Thomas Winyard}
\affiliation{School of Mathematics, University of Leeds, Leeds LS2 9JT, United Kingdom}
\author{Egor Babaev}
\affiliation{Department of Physics, KTH-Royal Institute of Technology, Stockholm, SE-10691 Sweden}

\begin{abstract}
We show that in superconductors that break time reversal symmetry and have anisotropy, such as $p+ip$ materials,
all order parameters and magnetic modes are mixed. Excitation of the gap fields produces an excitation of the magnetic field and vice versa.
Correspondingly the long-range decay of the magnetic field and order parameter are in general given by the same exponent.
Thus one cannot characterize $p+ip$ superconductors by the usual coherence and magnetic field penetration lengths.
Instead the system has normal modes that are associated with linear combinations of magnetic fields, moduli
of and phases of the order parameter components. Each such normal mode has its own decay length that plays
the role of a hybridized coherence/magnetic field penetration length. On a large part of the parameter space these exponents
are complex. Therefore the system in general has damped oscillatory decay of the magnetic field accompanied 
by damped oscillatory variation of the order parameter fields.

\end{abstract}

\maketitle

\section{Introduction}

Superconducting states that spontaneously break time reversal symmetry (BTRS)
are a subject of intense experimental pursuit.
Two types of BTRS state that attract particular interest are
chiral $p$-wave superconductors where the most intense discussions
were focused on $Sr_2RuO_4$ \cite{mackenzie2017ap,Mackenzie.Maeno:03},
and $s+is$ or $s+id$  superconducting states, evidence for which
was recently found in iron-based superconductors \cite{Grinenko2017,grinenko2018emerging}.
BTRS states are described by
an order parameter that has at least two components,
because they break at least $U(1)\times Z_2$ symmetry. Also, 
rather generically, there is anisotropy in such superconducting states.
In this work we investigate the most basic property of that state: the magnetic
field penetration and coherence lengths. 

The basic fundamental length scales of
superconductors were first discussed by Fritz and Heinz London \cite{London1935} and Ginzburg and Landau  \cite{landau1950k}
in an ordinary superconductor. This was done in the model for the simplest   superconductor  that breaks $U(1)$
symmetry, described by a single complex field $|\Psi|$ neglecting crystal anisotropies. The
London magnetic field penetration length $\lambda$ is the power in the exponential law of
decay of the magnetic field: ${\bf B}={\bf B}_0 e^{-r/\lambda}$.
The coherence length $\xi$ is the scale associated with the exponential law  describing how 
the modulus $|\Psi |$ of the complex field, describing the order parameter, restores  its
ground state value $\bar{|\Psi |}$ away from  a perturbation:
$|\Psi |{  (r)} \approx \bar{|\Psi |} - \rm{ const}  \ e^{ -r/\xi}$. 
The microscopic BCS theory of superconductivity
related the  modulus of the order parameter field $|\Psi |$ to a superconducting gap $\Delta$
in the single-electron spectrum.
The definition of the coherence length in the context
of superconductivity has an extra factor $\sqrt{2}$ which we absorb for brevity
in the definition of $\xi$.  
Often the coherence length is   assessed only approximately, and it is important to remember the limitations 
of these approximate definitions. For example while in the simplest Ginzburg-Landau model
coherence length is often estimated via  vortex core size or slope of the order parameter near the centre of the vortex core,
such estimates are known to fail even in the simplest models at low temperates \cite{Gygi1991}.
Another indirect way to assess coherence length assumes
its inverse proportionality to the gap function $\Delta$ in the BCS expression $\xi_0\propto 1/\Delta$.
Likewise   this expression  has very limited validity.
It cannot serve as an estimate  at strong coupling or in the multi-component case.
For example in the case of several gaps that would give unphysical divergence of coherence
length where a gap is closing (i.e.\ at the crossover from $s_{++}$ to $s_{\pm}$
states where   all coherence lengths should be finite because there is no symmetry breaking
and no accidental degeneracies). 
Similarly that estimate would miss the divergence of coherence
length when a superconductor transitions from ordinary to 
BTRS state i.e.  $s$ to $s+is$ or  $s$ to  $s+id$ state, whose existence is dictated by symmetry.
These examples shows that accurate coherence length calculations are required
while simple estimates can be highly misleading.
Calculations of coherence and magnetic field penetration  lengths have been made for isotropic multicomponent models for general 
interactions both in phenomenological and microscopic models \cite{Babaev.Carlstrom.ea:10,Carlstrom.Babaev.ea:11,silaev1,Carlstroem2011a,garaud2018properties}. 
The multi-component nature of these systems strongly affects only the coherence lengths, while 
the magnetic field penetration length is merely renormalized by intercomponent couplings.
The situation was found to be very different 
in $U(1)$ multiband superconductors if different bands have
different anisotropies. While usually the magnetic (London) modes decouple from
other normal modes of the system, such as density and phase difference (Leggett) modes,
having different  anisotropies in different bands results in a hybridization of the London mode  with
the phase difference mode  \cite{silaev2017non,winyard2018skyrmion,winyard2018}. For a system with $N$ bands that means that 
magnetic field decay is described by several modes with different exponents 
and there could be up $N+1$ such modes in the systems considered in \cite{silaev2017non,winyard2018skyrmion,winyard2018}.
Furthermore the powers in the corresponding exponents under certain conditions are complex leading to 
a damped oscillatory decay of the magnetic field.

That raises the question: what is the behaviour of the magnetic field and what are the coherence lengths in 
 $p+ip$ superconductors, since such systems are inherently both multicomponent and anisotropic?
The important difference with the systems considered in \cite{silaev2017non,winyard2018skyrmion,winyard2018}, as discussed
below, is the fact that such a
superconducting state has spontaneously broken time reversal symmetry.  

The standard Ginzburg-Landau model for a $p+ip$ superconductor can we written in dimensionless units as
\begin{equation}
{\cal F} = \frac12Q_{ij}^{\alpha\beta} D_i \psi_\alpha\overline{D_j \psi_\beta} + \frac12B^2 + F_p
\label{Eq:F}
\end{equation}
\noindent where the greek indices enumerate components of the order parameter and
latin indices stand for space directions.  
Summation over repeated indices is implied, $D_i = \partial_i - i A_i$ is the covariant derivative with the gauge field $A_i$, and the complex fields 
\begin{equation}
\psi_\alpha = \rho_\alpha e^{i\theta_\alpha} \ \  \ \alpha=1,2 
\end{equation}
represent the different superconducting components.  We consider here a quasi-two-dimensional system or a configuration of
a three dimensional system that   is translation invariant in the $z$ direction.
The magnetic field  $\boldsymbol{B} =(0,0,B)=(0,0,\cd_1A_2-\cd_2A_1)$ is directed so that the spatial indices take only the values $1,2$.
$F_p$ represents the potential terms which, by gauge invariance, may depend
only on $\rho_\alpha$ and $\theta_{12}:=\theta_1-\theta_2$. For the standard
$p+ip$ superconductor, the $\theta_{12}$ dependence enters only via a term of the form $(\psi_1\psi_2^*)^2+c.c.$, \cite{Heeb.Agterberg:99,Agterberg:98,Vadimov2013} that is,
\beq\label{thcu}
F_p=V(\rho_1,\rho_2)+\frac\eta8\rho_1^2\rho_2^2\cos{2\theta_{12}}
\eeq
with $\eta>0$. 
Then
the ground states  (minima) of $F_p$  are 
degenerate occurring with $\theta_{12}=\pm\pi/2$.  The ground state of this system is {not} gauge equivalent to its complex conjugate. Hence the system exhibits broken time reversal symmetry. Note that, although our focus below will be on
the example of $p+ip$ superconductors, the model is very general,
also describing other BTRS states such as $s+is$ and $s+id$ superconductors \cite{garaud2017microscopically,vadimov2018polarization}.
Our results obtained below apply also to such states when anisotropy is present.

The anisotropy of the system enters through the parameters $Q^{\alpha\beta}_{ij}$, which must satisfy
$Q^{\beta\alpha}_{ji}=(Q^{\alpha\beta}_{ij})^*$ to ensure $F$ is real. Henceforth we assume, as is standard, that all $Q^{\alpha\beta}_{ij}$ are real.

\section{Calculation of length scales}\label{sec:lin}
The spatial dependence of the fields at equilibrium is governed by the Ginzburg-Landau (Euler-Lagrange)  equations  for the functional $F=\int_{\R^2}{\cal F}$,
\bea
Q^{\alpha\beta}_{ij}D_iD_j\psi_\beta&=&2\frac{\cd F_p}{\cd \ol{\psi}_\alpha},\label{EL1}\\
\cd_j(\cd_jA_i-\cd_i A_j)&=&J_i,\label{EL2}
\eea
where the total supercurrent is
\beq
J_i:={\rm Im}(Q^{\alpha\beta}_{ij}\ol\psi_\alpha D_j\psi_\beta).
\eeq
 Consider the behaviour of the system a long distance from some defect (e.g.\ a vortex, domain wall or material boundary). Since the fields are close to their ground state values, they should be well approximated by solutions of the { linearization} of the Euler-Lagrange equations about the ground state. 
That is, since the characteristic exponents, such as coherence lengths, define the exponential decay of a
small perturbation of a field from its ground state, in order to calculate them one
expands fields in the Euler-Lagrange equations around their ground state values (see e.g. \cite{landau1980statistical,Plischke1989}).
For a conventional superconductor the coherence length is
obtained by expanding in small deviations of the field modulus $|\psi|$ \cite{Tinkham1995},
but that cannot a priori be done  for our system involving multiple fields.
Instead we should expand in small deviations in all degrees of freedom
and see if there is a coupling between the fields arising at the lowest order. 
Because we are dealing with a superconductor we have a coupling to the gauge field $A$ and some care must be taken in handling the gauge invariance of the system.  Let us define the phase field
\beq
\theta_{\Sigma}:=\frac12(\theta_1+\theta_2).
\eeq
 Note that $\rho_\alpha=|\psi_\alpha|$ and $\theta_{12}$ are gauge invariant, while $\theta_\Sigma$ and $A_i$ are not. The combination
\beq\label{pdef}
p_i:=A_i-\cd_i\theta_\Sigma
\eeq
{\em is} gauge invariant, and our strategy is to reexpress the Euler-Lagrange equations in terms of $\rho_\alpha$, $\theta_{12}$ and $p_i$. Let us denote the ground state values of $\rho_\alpha$ and $\theta_{12}$ by $u_\alpha$ and $\theta_0$ respectively; for the $p+ip$ model \eqref{thcu}, $\theta_0=\pm\frac\pi2$, but it is instructive to leave it general, for the time being.
Then saying that the fields are close to their ground state values means precisely that $p_i$, $\eps_\alpha$ and
$\theta_\Delta$ are small, where
\beq\label{edef}
\eps_\alpha:=\rho_\alpha-u_\alpha,\qquad \theta_\Delta:=\frac{1}{2}(\theta_{12}-\theta_0).
\eeq
In particular, the small quantities $\eps_\alpha,p_i,\theta_\Delta$ should obey the linearization of \eqref{EL2} about $(p_i,\rho_\alpha,\theta_{12})=(0,u_\alpha,\theta_0)$. The left hand side is exactly $\cd_j(\cd_jp_i-\cd_ip_j)$ which is already of linear order, but we must compute the supercurrent $J_i$ to linear order. This is straightforward once we recognize that $D_i\psi_\alpha$ is to linear order,
\bea
D_i\psi_1&=&(\cd_i\eps_1-i(p_i-\cd_i\theta_{\Delta})u_1)e^{i(\theta_\Sigma+\frac{1}{2}\theta_0)}+\cdots\nonumber\\
D_i\psi_2&=&(\cd_i\eps_2-i(p_i+\cd_i\theta_{\Delta})u_2)e^{i(\theta_\Sigma-\frac{1}{2}\theta_0)}+\cdots
\eea
so the linearization of \eqref{EL2} is
\bea
&&\cd_j(\cd_jp_i-\cd_i p_j)\nonumber \\&&=-Q^{11}_{ij}u_1^2(p_j-\cd_j\theta_\Delta)-Q^{22}_{ij}u_2^2(p_j+\cd_j\theta_\Delta)\nonumber \\
&&-u_1u_2\cos\theta_0\{Q^{12}_{ij}(p_j+\cd_j\theta_\Delta)-Q^{21}_{ij}(p_j-\cd_j\theta_\Delta)\}\nonumber\\
&&-\sin\theta_0\{Q^{12}_{ij}u_1\cd_j\eps_2-Q^{21}_{ij}u_2\cd_j\eps_1\},
\eea
Note that the left hand side of this equation is precisely the usual curl of the magnetic field ($p_i$ differs from $A_i$ by a gradient, so their curls coincide). The key observation is that, unless $\theta_0=0$ or $\pi$, that is, unless the ground state is phase locked or antilocked (or $Q^{12}_{ij}\equiv 0$) this PDE couples all    the   degrees of freedom together (through its final term), so that they all decay to zero with the same dominant length scale. Any other value of $\theta_0$ (including
$\pm\frac\pi2$) corresponds to a ground state $(\psi_1,\psi_2)=(u_1e^{i\frac{1}{2}\theta_0},u_2e^{-i\frac{1}{2}\theta_0})$ which is not gauge equivalent to its complex conjugate, and hence breaks time reversal symmetry. Hence, the effects described below are generic when one has BTRS and spatial anisotropy. 

To compute the length scales, we must linearize
\eqref{EL2} in $(p_i,\eps_\alpha,\theta_\Delta)$ also. Henceforth, we specialize to the $p+ip$ case with potential \eqref{thcu}, so that $\theta_0=\pm\frac\pi2$. 
Substituting \eqref{pdef},\eqref{edef} into the Euler-Lagrange equations and discarding all terms nonlinear in small quantities yields
\begin{eqnarray}
\nonumber -\left(\partial_1^2 + \partial_2^2\right)p_i + \partial_i\partial_j p_j - L_{ij} \partial_j \theta_\Delta + K_{ij} p_j &&\\ \pm Q^{12}_{ij}\left(u_1\partial_j \eps_2 - u_2 \partial_j \eps_1\right) &=& 0\label{Eq:eom_p}\\
\nonumber \pm Q^{12}_{ij}\left(u_2 \partial_i \partial_j \eps_1 + u_1 \partial_i \partial_j \eps_2\right) - K_{ij}\partial_i \partial_j \theta_\Delta && \\+ L_{ij} \partial_i p_j + 2\eta u_1^2 u_2^2 \theta_\Delta&=&0\label{Eq:eom_t}\\
-Q^{11}_{ij} \partial_i \partial_j \eps_1 \pm Q^{12}_{ij}u_2 \partial_i\left(p_j + \partial_j\theta_\Delta\right) + \mathcal{H}_{1\beta} \eps_\beta&=&0\label{Eq:eom_e1}\\
-Q^{22}_{ij} \partial_i \partial_j \eps_2 \mp Q^{12}_{ij}u_1 \partial_i\left(p_j - \partial_j\theta_\Delta\right) + \mathcal{H}_{2\beta} \eps_\beta&=&0\label{Eq:eom_e2}
\end{eqnarray}
where we have defined the matrix coefficients,
\begin{eqnarray}
K_{ij} = Q^{11}_{ij} u_1^2 + Q^{22}_{ij} u_2^2,\\
L_{ij} = Q^{11}_{ij} u_1^2 - Q^{22}_{ij} u_2^2,
\end{eqnarray}
and
\beq
\mathcal{H}_{\alpha\beta} = \left.\frac{\partial^2 F_p}{\cd\rho_\alpha\cd\rho_\beta}\right|_{\left(u_1,u_2, \pm \frac{\pi}{2}\right)}
\eeq
is the Hessian of the potential $F_p$ about the ground state, with respect to $(\rho_1,\rho_2)$.

Note that in the case where there are no
mixed gradient terms $Q^{12}=0$ the linearized equations decouple into a pair for $(p_i,\theta_\Delta)$ and a pair for $(\eps_1,\eps_2)$.
That means that small fluctuations in the density fields do not cause a perturbation
of the phase difference and do not create magnetic field, as is indeed the case in ordinary superconductors,
or in the class of anisotropic models studied in \cite{silaev2017non,winyard2018,winyard2018skyrmion}.
However we see that for anisotropic superconductors that break time reversal symmetry,
 such as $p+ip$ superconductors,
no such simplification takes place: all the gauge invariant fields $\eps_\alpha, \theta_\Delta, p_i$ are coupled to one another, and 
when one changes all the others should change too.
The implication of this is that
systems like chiral $p+ip$ superconductors {\it cannot be characterized by coherence and magnetic
field penetration lengths in the usual sense, but the decay length scale of a small perturbation of the order parameter field and magnetic field is in general the same.}
Furthermore it implies that one cannot reliably use the London limit to calculate the magnetic field penetration length because density modes do not asymptotically decouple from magnetic modes.
Below we calculate these length scales.

 Since the equations are anisotropic, to extract the length scales we must first select a direction (normal to the domain wall or material boundary, or radial from the vortex core, depending on context), denoted by a unit vector $\nv=(n_1,n_2)$, and then reduce the equations to  ordinary differential equations (ODEs)  with $\nv$-dependent coefficients, by imposing translation invariance orthogonal to $\nv$. So, we demand that
\begin{eqnarray}
p_i &=& a(X)n^\perp_i + b(X)n_i \nonumber \\
\theta_\Delta &=& \theta_\Delta(X),\qquad \eps_\alpha = \eps_\alpha(X)\label{ansatz}
\end{eqnarray}
where $X = n_i x_i$ and $\nv^\perp=(-n_2,n_1)$. Substituting \eqref{ansatz} into \eqref{Eq:eom_p}-\eqref{Eq:eom_e2},
one obtains a coupled set of five ODEs. 
The two-vector valued ODE (\ref{Eq:eom_p}) implies a pair of scalar-valued ODEs, obtained by taking its scalar product with $\nv$ and $\nv^\perp$. 
The $\nv$ component implies
\begin{equation}
b = \frac{-\nv}{\nv\cdot K\nv}\cdot\left(K\nv^\perp\, a -L\nv\, \theta_\Delta' \pm Q^{12}\nv(u_1\eps_2'-u_2\eps_1')\right)
\end{equation}
(where ${}'\equiv d/dX$),
which can be used to eliminate $b(X)$ from the other ODEs.
We now have 4 coupled ODEs, forming a linear system, that describes the response of the system to a small perturbation about its ground state.
\begin{equation}
\AA \vec{w}'' + \BB \vec{w}' + \CC\vec{w} = 0,
\label{Eq:polynomial}
\end{equation}
where $\vec{w} = \left(\eps_1, \eps_2,\theta_{\Delta},a\right)^T$ and $\AA$, $\BB$, $\CC$ are certain constant $4\times 4$ real matrices. It is important to note that $\AA$ and $\CC$ are symmetric, while $\BB$ is skew, and that all three depend on the choice of direction $\nvec$. Their exact form is given in Appendix A. 

Recall that \eqref{Eq:polynomial} is the linearized system of field equations describing how  a system recovers
from a perturbation in the $\nv$-direction under the assumption of translation invariance orthogonal to $\nv$,
for example, how the system behaves near the boundary of a superconductor subject to an external magnetic field.
Its general solution is
\beq\label{gensol}
\vec{w}(X)=\sum_{i=1}^8c_i\vec{v}_ie^{-\mu_iX}
\eeq
where $\mu_1,\mu_2,\ldots,\mu_8$, are the  solutions of the degree 8 polynomial equation
\begin{equation}
\det\left(\mu^2 \AA - \mu \BB + \CC\right) = 0.
\label{Eq:kernel}
\end{equation}
The constants $\mu_i$ should be interpreted as field masses which set the length scale $\lambda_i$ of spatial  decay of the associated  linear combination of
fields via
\begin{equation} 
\lambda_i=\frac{1}{\mu_i}.
\end{equation}
The quantities 
$\vec{v}_1,\vec{v}_2,\ldots,\vec{v}_8$, are the corresponding eigenvectors (by eigenvector we mean a unit length vector satisfying $(\mu_i^2 \AA - \mu_i \BB + \CC)\vec{v}_i=\vec{0}$), and $c_1,c_2,\ldots,c_8$ are  arbitrary constants, determined by boundary conditions and nonlinearities.
Each exponential power is associated to a normal mode, determined by $\vec{v}_i$. In an ordinary superconductor the 
normal mode associated with the coherence length is the modulus of the order parameter, while
the magnetic field penetration length is attributed to a massive vector field: the magnetic field.
Instead, we see that in the chiral $p+ip$ superconductor the normal modes are
associated with linear combinations of magnetic and matter degrees of freedom.
 
Indeed, the polynomial equation (\ref{Eq:kernel}) has real coefficients and is quartic in $\mu^2$ (since $\AA,\CC$ are symmetric, while $\BB$ is skew);
hence, if $\mu$ is a solution, so are $-\mu$, $\mu^*$ and $-\mu^*$.
This demonstrates that complex length scales are caused by mixing, as this is the only way for multiple length scales to become linked and hence be complex conjugates of each other. Exactly half the eigenvalues, which we choose to label $\mu_1,\ldots,\mu_4$ have positive real part, while the others have negative real part. We seek solutions that decay to $0$ as $X\ra\infty$; these are obtained by setting $c_i=0$ for $i\geq 5$ in
equation (\ref{gensol}). 

The long-range behaviour of the fields, in direction $\nvec$, is governed by the dominant eigenvector $\vec{v}_i=\vec{v}_*$, defined to be the eigenvector whose eigenvalue $\mu_i=\mu_*$ has smallest positive real part (hence the
longest length scale $\lambda_*=1/\mu_*$ of spatial decay). Note that, in general, $\mu_*$ may be {\em complex}, in which case the fields at large $X$ are spatially oscillatory, behaving like
\beq\label{offa}
(\eps_1,\eps_2,\theta_{\Delta},a)\sim  c\vec{v}_{r,*} e^{-\Re(\mu_*) X}\cos\Im(\mu_*) X,  \ \ \ (X \to \infty)
\eeq
where $c$ is some real constant and $\vec{v}_{r,*}$ is the real part of $\vec{v}_*$. 
 The complex magnetic field penetration length implies oscilatory decay of the magnetic field as observed in anisotropic systems without BTRS \cite{silaev2017non,winyard2018,winyard2018skyrmion}.  Here we find that in  a $p+ip$ superconductor one cannot assume that a perturbation of the gap fields will decay with real exponents: i.e.\ there are no real coherence lengths in general.
Note that for dirty isotropic multiband superconductors, the phase difference and density modes can be mixed even without breaking time reversal symmetry \cite{garaud2018properties}. Our findings of the complete mixing of the order parameters and magnetic modes would apply also for that case.

 Importantly, as detailed below, in general one needs to retain contributions from the modes associated with shorter length scales.

Of course, our analysis should reproduce the usual picture of separate real length scales (the coherence length and magnetic penetration depth) in the case of a spatially isotropic system, where $Q^{\alpha\beta}_{ij}=\delta_{\alpha\beta}\delta_{ij}$, and should hold approximately for a small perturbation of this.
In the near-isotropic regime, when
$Q^{11},Q^{22}\approx I_2$ and $Q^{12}\approx 0$, the coupling between $\eps_\alpha,\theta_\Delta,a$ is weak,
the spectrum is real, and one of the eigenvectors, $\vec{v}_{4}$ say, is approximately $(0,0,0,1)$, while the others, $\vec{v}_1,\vec{v}_2,\vec{v}_3$, are approximately normal to $(0,0,0,1)$. We then recover the usual picture of separate length scales associated with the magnetic field, $\lambda_{mag}=\lambda_4$, and the condensates, $\lambda_{1,2,3}$. Consider the case where $\lambda_*=\lambda_{mag}$, that is, $\vec{v}_{4}$ is dominant. Although all fields do, strictly speaking, decay like \eqref{offa} (with $\Im(\lambda_*)=0$) at very large $X$, the coefficients in front of $\eps_1,\eps_2,\theta_\Delta$ are very small, while the coefficient 
in front of $a$ is of order of unity ($\vec{v}_*=\vec{v}_4\approx(0,0,0,1)$) so at intermediate range contributions from the subdominant eigenvectors are larger. This  allows one to identify approximately  $\lambda_{mag}$ as a penetration depth and $\min\{\lambda_i\}$ as a coherence length, and classify the system as
type-2 (since $\lambda_{mag}$ is the largest length scale). Similar remarks apply if one of the condensate modes is dominant. 
This is consistent with the numerical solutions obtained earlier in such regimes \cite{Garaud2016a}.
One therefore may approximately call the exponents associated to matter-field-dominated modes coherence lengths and 
those associated with magnetic-field dominated modes magnetic field penetration lengths.
However this approximate picture disappears as one increases the anisotropy and 
magnetic and matter field   couplings in \eqref{Eq:polynomial} become significant. 

In summary, the long range behaviour of spatially decaying solutions of our system is \eqref{offa} where $\mu_*=1/\lambda_*$ is the 
solution of \eqref{Eq:kernel} with smallest positive real part. In general, $\mu_*$ depends on $\nvec$, the direction along which we impose spatial decay, and may be complex, in which case the decay of both magnetic and gap fields is oscillatory. We have also shown that this coupling and the oscillations in all four fields is a direct result of BTRS and anisotropy.

In the next section we consider the implications of these findings for the Meissner state
of a   $p+ip$ superconductor.

\section{Meissner State in a $p+ip$ System}

We consider   a simple   $p+ip$ model, such as the one discussed in the context of  the debate of 
the nature of superconducting state in Sr${}_2$RuO${}_4$    in \cite{bouhon2010influence}. This is of the form
\eqref{Eq:F} with (after a trivial rescaling of fields which is shown in detail in appendix D)
\begin{gather*}
Q^{11} = \left(\begin{array}{cc} 3 + \nu & 0 \\ 0& 1-\nu\end{array}\right), \quad Q^{22} = \left(\begin{array}{cc} 1 - \nu & 0 \\ 0& 3 + \nu\end{array}\right), \\ \quad Q^{12} = \left(\begin{array}{cc} 0 & 1-\nu \\ 1-\nu & 0\end{array}\right)
\label{Eq:aniBS}
\end{gather*}
and potential,
\begin{eqnarray}
&F_p& = V_0\left\{1 -(\rho_1^2+\rho_2^2)+ \frac{1}{8}\left(3+\nu\right)\left(\rho_1^2 + \rho_2^2\right)^2 \right. \nonumber \\  &-& \left. \frac{1}{4}\left(1+3\nu\right)\rho_1^2\rho_2^2  + \frac{1}{4}\left(1 - \nu\right)\rho_1^2\rho_2^2\cos{2\theta_{12}}\right\}.
\label{Eq:FpBS}
\end{eqnarray}
The model contains two unknown parameters: $-1<\nu<1$ which measures the anisotropy of the Fermi surface, and $V_0$, the overall strength of the potential (coinciding with $b/(\pi\gamma^2K^2)$ in the  notation of Ref. \cite{bouhon2010influence}). 
Its ground states are $(\rho_\alpha, \theta_{12}) = (1,\pm \pi/2)$, so $u_1=u_2=1$.

\begin{figure*}
  \centerline{\fbox{\includegraphics[width=0.92\linewidth]{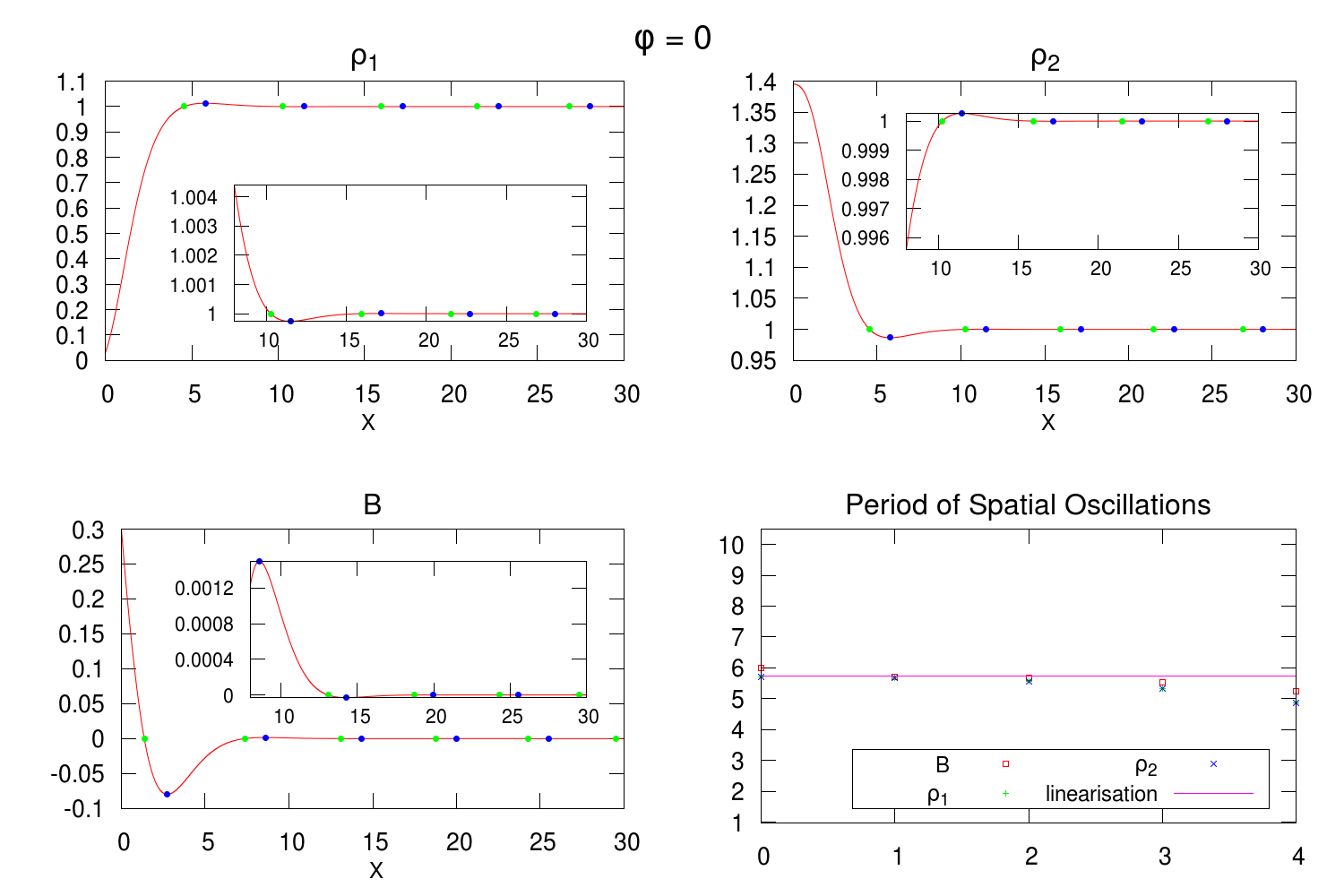}}}
  \centerline{\fbox{\includegraphics[width=0.92\linewidth]{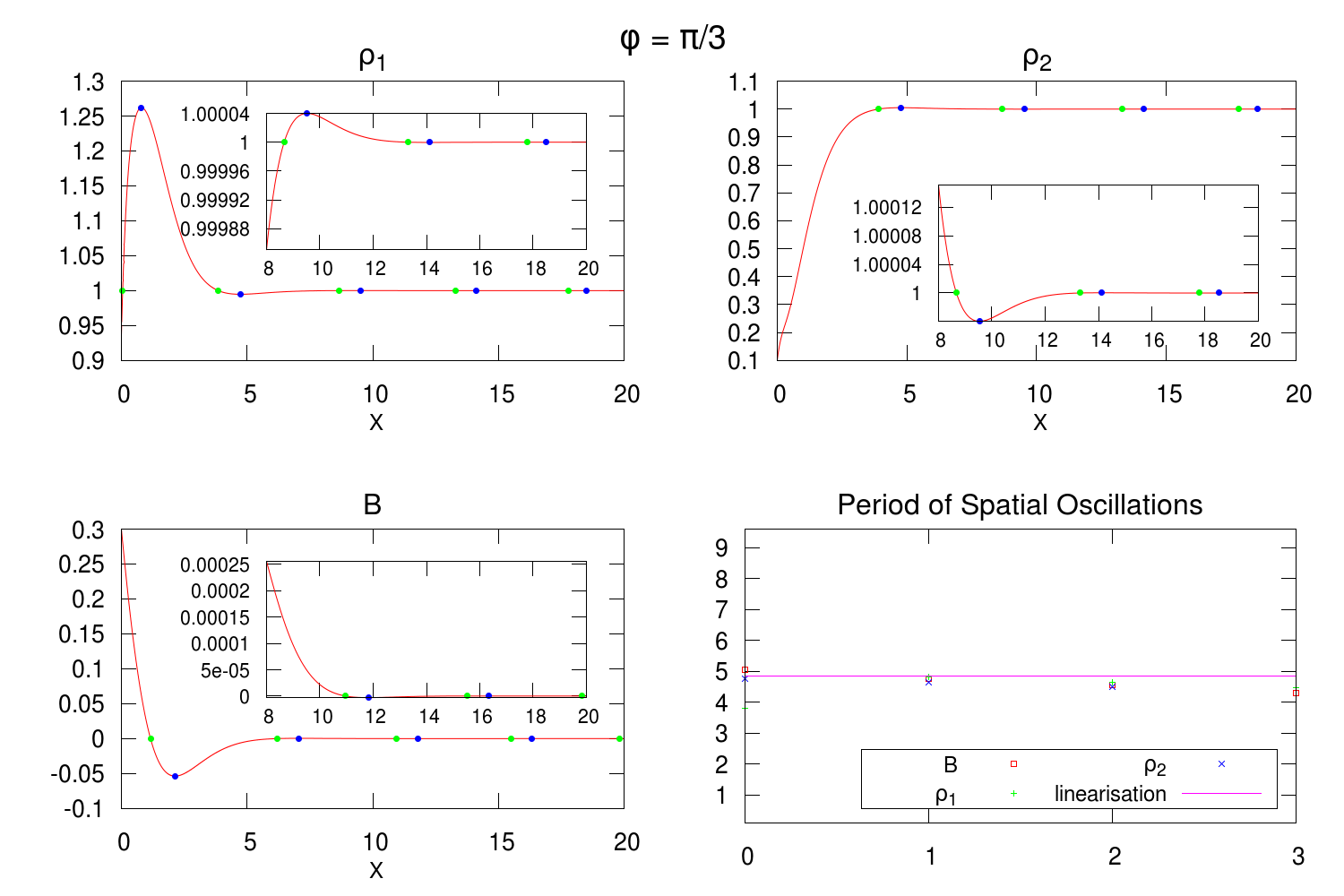}}}
 \caption{\label{Fig:Meissner1} 
Superconductor-insulator boundary of a $p+ip$ superconductor with $V_0 = 3$, $\nu = -0.95$, $\chi=1$ and external field $H=0.3$ for two different boundary orientations: $\varphi=0$ (top set of four plots) and $\varphi=\pi/3$ (bottom set of four plots). The boundary is at $X=0$, the plotted fields are the condensate magnitudes $\rho_1$ and $\rho_2$ and the magnetic field strength $B$. The green dots mark points where the spatially oscillating fields cross their ground state values and the blue dots mark local extrema. The distances between these successive points are compared with the prediction of our linear analysis in the bottom right plot of each set. 
}
\end{figure*}

\begin{figure*}
  \centerline{\includegraphics[width=1.05\linewidth]{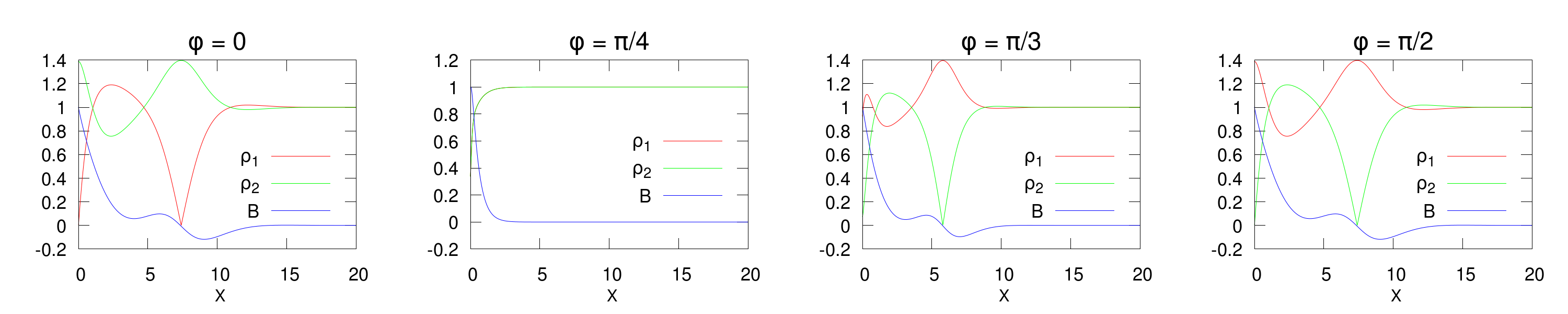}}
 \caption{\label{Fig:Meissnerzero} 
Superconductor-insulator boundary at high external field $H = 1$ and boundary orientations $\varphi = 0,\frac{\pi}{4}, \frac{\pi}{3}, \frac{\pi}{2}$, showing
stripe formation in the Meissner state: one condensate component goes to zero and the other achieves a maximum, producing a stripe (orthogonal to $\nv$) of depletion of one condensate and surfeit of the other. Note that the $\rho_1$ and $\rho_2$ curves coincide in the case $\varphi=\frac\pi4$. The model parameters are as in Figure \ref{Fig:Meissner1}). }
\end{figure*}

Consider a semi-infinite superconductor occupying the half-space $X\geq 0$ (where, as before,
$X=n_1x_1+x_2n_2$), denoted $\Omega$, with the region $X<0$ occupied by an insulator. Denote by $\cd\Omega$ the boundary between these regions (where $X=0$). Note that $\nv=(\cos\vphi,\sin\vphi)$ is an inward pointing unit normal to this boundary. The system is subjected to a uniform external magnetic field $H$ in the $x_3$ direction. Provided $H$ is not too strong, the system will approach the ground state $\rho_1=\rho_2=1$, $\theta_{12}=\pi/2$ (say) in the bulk (as $X\ra \infty$). 
To find the Meissner state, we minimize the Gibbs free energy
\beq
G=\int_\Omega{\cal F}-H\int_{\Omega}B+\int_{\cd\Omega}{\cal F}_{\mbox{surf}}
\label{Eq:Gibbs}
\eeq
over all fields in $\Omega$, assuming invariance under translations normal to $\nv$.
Here, we use the standard boundary conditions, advocated in \cite{RevModPhys.63.239}, by including in the free energy the surface term
\begin{eqnarray}
{\cal F}_{\mbox{surf}}&=& \chi_1(\rho_1^2+\rho_2^2)+\chi_2(n_1^2-n_2^2)(\rho_1^2-\rho_2^2)\nonumber\\& &+2\chi_3n_1n_2(\psi_1^*\psi_2+\psi_1\psi_2^*)
\label{Eq:surf}
\end{eqnarray}
For simplicity, we assume reflection from the boundary is {\em specular}, meaning that $\chi_1=\chi_2=\chi_3=\chi>0$. Having imposed translation invariance, the problem
reduces to a one-dimensional variational problem on $[0,\infty)$, with natural boundary conditions at $0$, which can be solved by a standard gradient-descent method. A more detailed discussion of the boundary conditions is given in appendix B. There is a caveat here. It has been demonstrated recently for $s$-wave superconductors, that boundary conditions can be different in superconductors from those based on the standard assumptions of Caroli-deGennes-Matricon type theory \cite{samoilenka}, which implies that the standard theory of boundary conditions for $p+ip$ should also be revised. However here we are interested not in the precise field values at the boundary but rather in
the laws governing their decay away from the boundary. Therefore the precise form of the boundary conditions is not very important. In Appendix C we present results with the extra boundary terms omitted entirely, giving the same field decay behaviour. 

The solutions depend on the unknown model parameters
$\nu,V_0,\chi$ as well as the applied field $H$ and the boundary orientation angle $\vphi$. We have run simulations for $\chi\in\{0,0.01,0.1,1,10\}$, finding no qualitative change in the physics we are focussed on. For that reason we fix $\chi = 1$ for the remainder of this section and present a representative sample of the other
parameters. We have included a plot, figure \ref{Fig:Meissner1xi0} for $\chi = 0$ in Appendix C for comparison, to demonstrate that the oscillatory behaviour of the fields originates in complex coherence lengths, not from the 
boundary terms in eq.\ (\ref{Eq:surf}).

For $V_0 = 3$, $\nu = -0.95$, $H = 0.3$, the Meissner states with boundary orientations $\vphi=0$ and $\vphi=\pi/3$ are presented in figure \ref{Fig:Meissner1}. Both exhibit oscillatory tails and field inversion of both $B$ and the condensates, consistent with exponential decay with a complex coherence and magnetic field penetration lengths. We shall return to this shortly. 
If the external field $H$ is increased further, the condensates separate more until, for some $\varphi$, one or other of the densities $\rho_1$ or $\rho_2$ hits zero. This produces a novel Meissner state depicted in figure \ref{Fig:Meissnerzero} for $H=1$
(well below the lower critical field $H_{c1}=1.34$) for several angles $\varphi$. We see that neither matter field component vanishes for $\varphi = \pi/4$, whereas $\rho_1$ vanishes for $\varphi = 0$, and $\rho_2$ vanishes for $\varphi = \pi/2$. As one  condensate component goes to zero, the other achieves a maximum exceeding its ground state value, producing a stripe (orthogonal to $\nv$) of depletion of one condensate and surfeit of the other.

\begin{figure}
  \centerline{\includegraphics[width=0.47\linewidth]{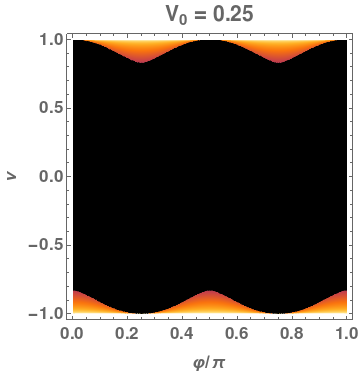}\includegraphics[width=0.47\linewidth]{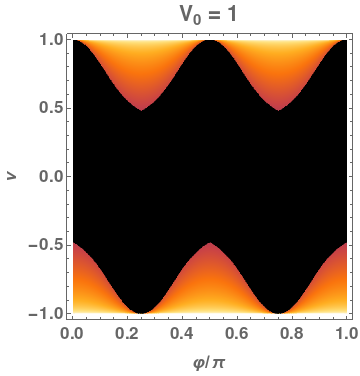}\includegraphics[trim={0.0cm -0.28cm 0 0},clip,width=0.095\linewidth]{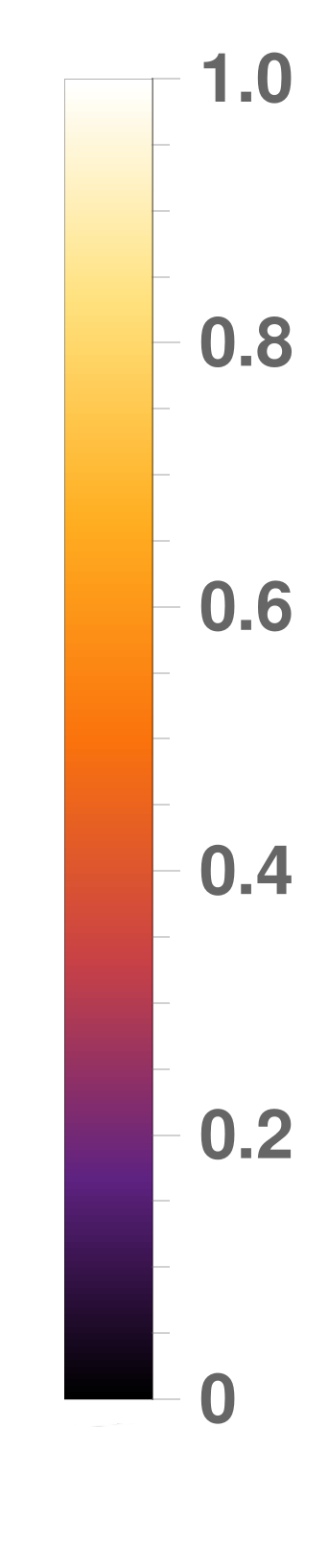}}\centerline{\includegraphics[width=0.47\linewidth]{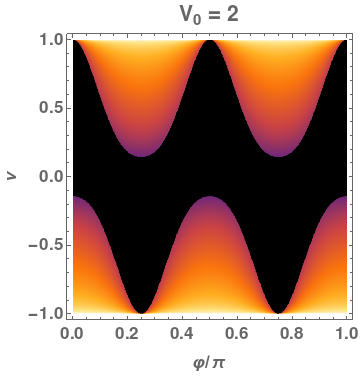}\includegraphics[width=0.47\linewidth]{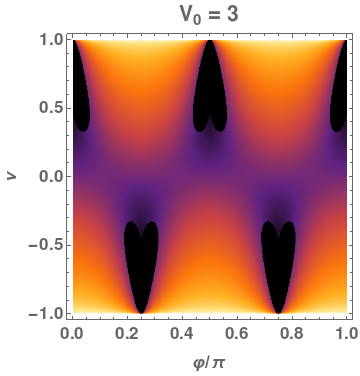}\includegraphics[trim={0.0cm -0.28cm 0 0},clip,width=0.095\linewidth]{Leadingbar2}}
 \caption{\label{Fig:lambdastar} Plots of $\left|Im\left(\mu_\star\right)\right|/\left|Re\left(\mu_\star\right)\right|$, where $\mu_\star=\lambda_*^{-1}$ is the leading mass scale (inverse length scale with smallest real part), in the $(\vphi,\nu)$ parameter space, for various values of $V_0$. Here $\vphi$ is the orientation of the sample boundary, and $\nu$, $V_0$ are parameters in the GL energy controlling the spatial anisotropy and the potential energy scale respectively. The black regions indicate where $\mu_\star$ is real and hence there will be no oscillations of the magnetic field, or condensates, away from the sample boundary.}
\end{figure}

Returning to our main goal of testing the analysis of the previous section, 
it is straightforward to compute, for any given $\vphi$ (boundary orientation), $\nu$ (anisotropy parameter) and $V_0$ (potential energy scale in the GL energy) the dominant eigenvalue $\mu_*$, and hence
map out the parameter set on which $\mu_*$ is complex. Figure \ref{Fig:lambdastar} presents pictures of the $(\vphi,\nu)$ parameter plane, for a sequence of values of $V_0$, coloured to show the parameter domain where $\mu_*$ is complex. For $V_0$ small, the parameter domain of complex $\mu_*$ is small and confined to the edges where $|\nu|$ is close to $1$, but as $V_0$ increases, the domain swells, eventually covering the whole parameter space (when $V_0\approx 4$), predicting that the Meissner state should be spatially oscillatory for all anisotropies $\nu$ and all boundary orientations $\vphi$ if $V_0$ is around this value. 
Increasing $V_0$ still further, pockets of real $\mu_*$ return and gradually refill the whole parameter space for very high values of $V_0$.
Turning to the parameter sets of figure \ref{Fig:Meissner1}, $V_0=3$, $\nu=-0.95$ and $\vphi=0,\pi/3$, we find in both cases that $\mu_*$ is complex, consistent with the nonlinear numerics 
($\mu_*(0)=0.689 + 0.548 i$, $\mu_*(\pi/3)=1.013 + 0.648 i$).
 The oscillatory decay predicted by linear analysis predicts that the zeros of
$B$, and of $\rho_\alpha-u_\alpha$ should be equally spaced with period $\pi/\Im(\mu_*)$, as should successive extrema of these functions. These gap widths can easily be extracted from the nonlinear numerics, and
are displayed, for these parameter sets, alongside the linear prediction, in figure \ref{Fig:Meissner1}. The agreement is remarkable. Finally, we have also chosen parameter sets for which $\mu_*$ is real, so that the linearization predicts non-oscillatory decay.
While the solutions still generically exhibit a single peak in each field, after this initial overshoot in the nonlinear regime, the fields decay exponentially without oscillation as the linearization predicts. 

\section{Conclusions}

In conclusion we have shown that the normal modes in an anisotropic superconductor that breaks time reversal symmetry mix density and phase fields with the magnetic field. This precludes using the usual notion of coherence and magnetic field penetration lengths because long range decay of matter and magnetic fields is given by the same exponent. Additionally the fundamental length scales,   associated with the normal modes, that mix the order parameters and magnetic field, are in general complex. We have also shown that this mode mixing requires BTRS along with anisotropy and that mixing is required for complex length scales. While systems exist with oscillations in magnetic field and phase difference due to anisotropy driven mixing between these two modes \cite{silaev2017non}, all four fields having complex length scales can only happen in an anisotropic BTRS system.
Calculating numerically the Meissner effect in a chiral $p+ip$ superconductor, we indeed find that application of an external
magnetic field is screened in an oscillatory way and produces damped oscillatory decay of the order parameter fields. For strong anisotropy the effect should be detectable in muon spin relaxation experiments. Cutting sample boundaries under different angles relative to crystal axes and measuring magnetic field inversion can allow one to recover information about the order parameter. Finally we note that our analysis dictates that the effect is present for any inhomogeneous situation, including the domain wall excitations in $p+ip$ superconductors considered in \cite{bouhon2010influence}; these should also exhibit oscillation and field inversion. That this was not observed in \cite{bouhon2010influence} might be an artifact of an overly restrictive ansatz. We plan to examine this further in a separate publication.

\appendix
\section{A: The coupling matrices}
Here we record the non-zero matrix elements of the $4\times 4$ matrices appearing in equation \ref{Eq:polynomial}:
\begin{eqnarray}
\AA_{11} &=& -n\cdot Q^{11} n + \frac{\left(n \cdot Q^{12} n\right)^2}{n\cdot K n}u_2^2,\\
\AA_{12} &=& -\frac{\left(n\cdot Q^{12} n\right)^2}{n\cdot K n}u_1 u_2, \\
\AA_{13} &=& \pm u_2 n\cdot Q^{12}n\left(1+\frac{n\cdot L n}{n \cdot K n}\right),\\
\AA_{22} &=& -n \cdot Q^{22} n + \frac{\left(n\cdot Q^{12} n \right)^2}{n \cdot K n} u_1^2, \\
\AA_{23} &=& \pm u_1 n\cdot Q^{12}n\left(1 - \frac{n \cdot L n}{n \cdot K n}\right),\\
\AA_{33} &=& \frac{\left(n \cdot L n\right)^2}{n \cdot K n} - n\cdot K n,\\
\AA_{44} &=& -1 \\
\BB_{14} &=& \pm u_2 \left(n\cdot Q^{12} n^\perp - n \cdot Q^{12} n \frac{n \cdot K n^\perp}{n \cdot K n}\right),\\
\BB_{24} &=& \mp u_1 \left( n\cdot Q^{12} n^\perp - n \cdot Q^{12} n \frac{n \cdot K n^\perp}{n\cdot K n}\right), \\
\BB_{34} &=& \left( n\cdot L n^\perp - n\cdot K n^\perp \frac{n\cdot L n}{n \cdot K n}\right),\\
\CC_{\alpha\beta} &=& {\cal H}_{\alpha\beta},\quad 1\leq\alpha,\beta\leq 2\\
\CC_{33} &=& 2\eta u_1^2 u_2^2,\\
\CC_{44} &=& n^\perp \cdot K n^\perp - \frac{\left(n\cdot K n^\perp\right)^2}{n\cdot K n}.
\end{eqnarray}
Recall that $\AA_{ij}\equiv \AA_{ji}$, $\BB_{ij}\equiv -\BB_{ji}$ and $\CC_{ij}=\CC_{ji}$. 

\section{B: Boundary Conditions}

To compute the Meissner state in the region $\Omega$ numerically we must minimize the Gibbs free energy
\beq
G=\int_{\Omega}({\cal F}-HB)+\int_{\cd\Omega}{\cal F}_{\mbox{surf}}=:\int_\Omega{\cal G}+\int_{\cd\Omega}{\cal F}_{\mbox{surf}}
\eeq
among all fields defined on $\Omega$. It is convenient to include a gauge-fixing term $\frac12(\cd_i A_i)^2$ in
${\cal F}$, and to denote the dynamical fields collectively as $\phi_a$, $a=1,\ldots,6$ (consisting of the real and imaginary parts of $\psi_\alpha$, and $A_1$, $A_2$). Then, under a variation $\delta\phi_a$, $G$ varies as
\begin{align}
\delta G=&\int_{\Omega}\left(
\frac{\cd {\cal G}}{\cd\phi_a}-\cd_i\left(
\frac{\cd{\cal G}}{\cd(\cd_i\phi_a)}\right)
\right)\delta\phi_a\nonumber\\
&+
\int_{\cd\Omega}\left(\frac{\cd {\cal F}_{\mbox{surf}}}{\cd\phi_a}-n_i\frac{\cd{\cal G}}{\cd(\cd_i\phi_a)}\right)\delta\phi_a,
\end{align}
where we have used the divergence theorem, and recalled that $\nv$ is an {\em inward} pointing normal to $\cd\Omega$. Demanding that $\delta G=0$ for all variations requires both these integrals vanish identically, and hence that $\phi_a$ satisfy the usual Euler-Lagrange equations in $\Omega$ together with the boundary conditions
\beq
\frac{\cd {\cal F}_{\mbox{surf}}}{\cd\phi_a}-n_i\frac{\cd{\cal G}}{\cd(\cd_i\phi_a)}=0
\eeq
on $\cd\Omega$. For the model studied here, this reduces to
\bea
n_iQ^{1\beta}_{ij}D_j\psi_\beta&=&2[(\chi_1+\chi_2(n_1^2-n_2^2))\psi_1+2\chi_3n_1n_2\psi_2],\nonumber\\
n_iQ^{2\beta}_{ij}D_j\psi_\beta&=&2[(\chi_1-\chi_2(n_1^2-n_2^2))\psi_2+2\chi_3n_1n_2\psi_1],\nonumber\\
\cd_iA_i&=&0,\nonumber \\
B&=&H.
\eea
Imposing the translationally invariant ansatz $\psi_\alpha=\psi_\alpha(X)$, $A_i=a(X)n_i^\perp+b(X)n_i$,
where $X=n_ix_i$, this reduces further to
\begin{align}
\nvec\cdot Q^{1\beta}\nvec&(\psi_\beta'(0)+ib(0)\psi_\beta(0))+i\nvec\cdot Q^{1\beta}\nvec^\perp a(0)\psi_\beta(0)
\nonumber\\ &=2[(\chi_1+\chi_2(n_1^2-n_2^2))\psi_1(0)+2\chi_3n_1n_2\psi_2(0)],\nonumber\\
\nvec\cdot Q^{2\beta}\nvec&(\psi_\beta'(0)+ib(0)\psi_\beta(0))+i\nvec\cdot Q^{2\beta}\nvec^\perp a(0)\psi_\beta(0)
\nonumber\\ &=2[(\chi_1-\chi_2(n_1^2-n_2^2))\psi_2(0)+2\chi_3n_1n_2\psi_1(0)],\nonumber\\
 b'(0)&=0, \nonumber\\
 a'(0)&=H.
\end{align}
These are the boundary conditions we impose at $X=0$. At $X=L$, large (our effective infinity), we demand that
$b'=a'=0$, $\psi_1=u_1$ and $\psi_2=iu_2$ (the fields are in their ground state state).

\section{C: $\chi = 0$ results}
\begin{figure*}
  \centerline{\fbox{\includegraphics[width=0.92\linewidth]{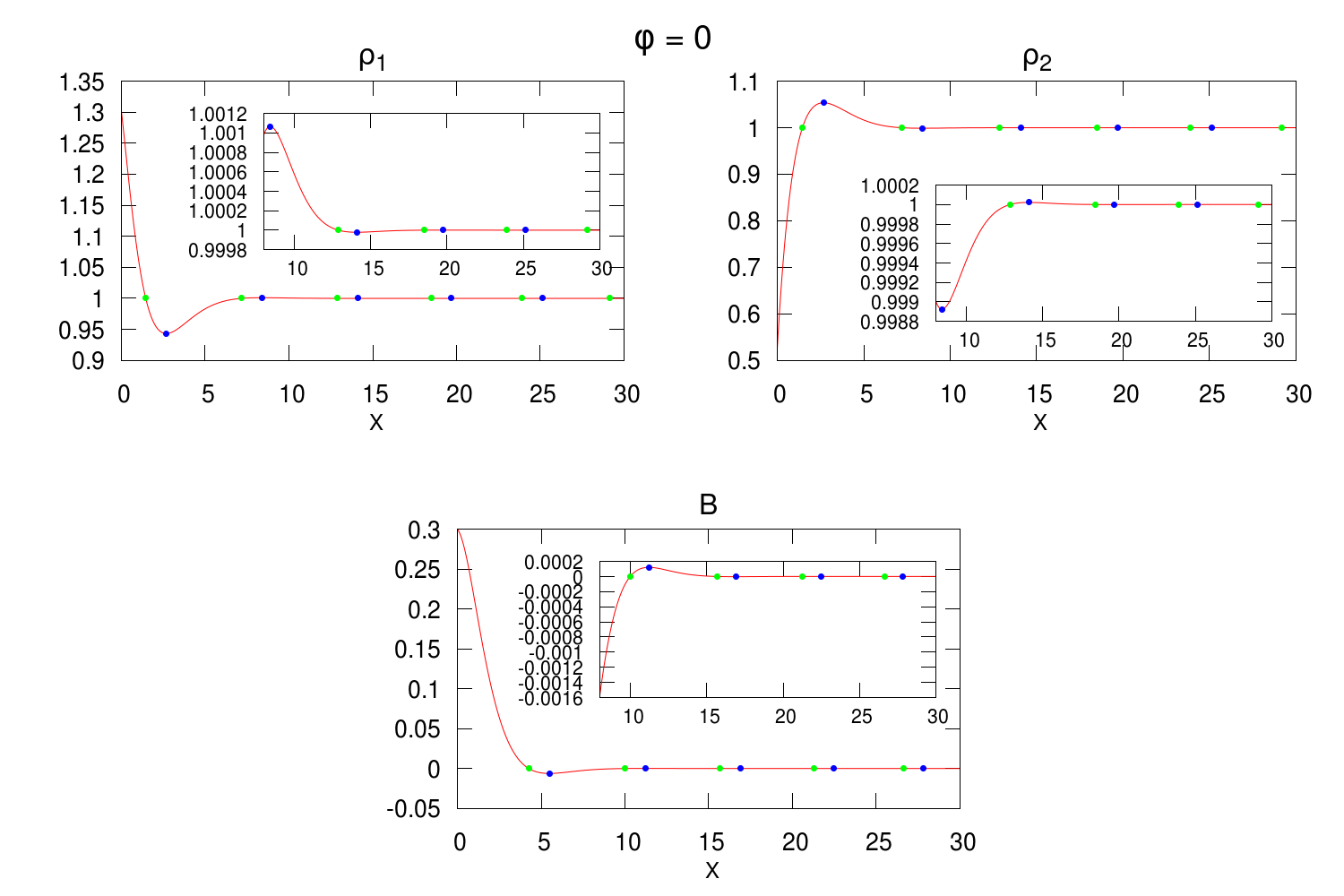}}}
  \centerline{\fbox{\includegraphics[width=0.92\linewidth]{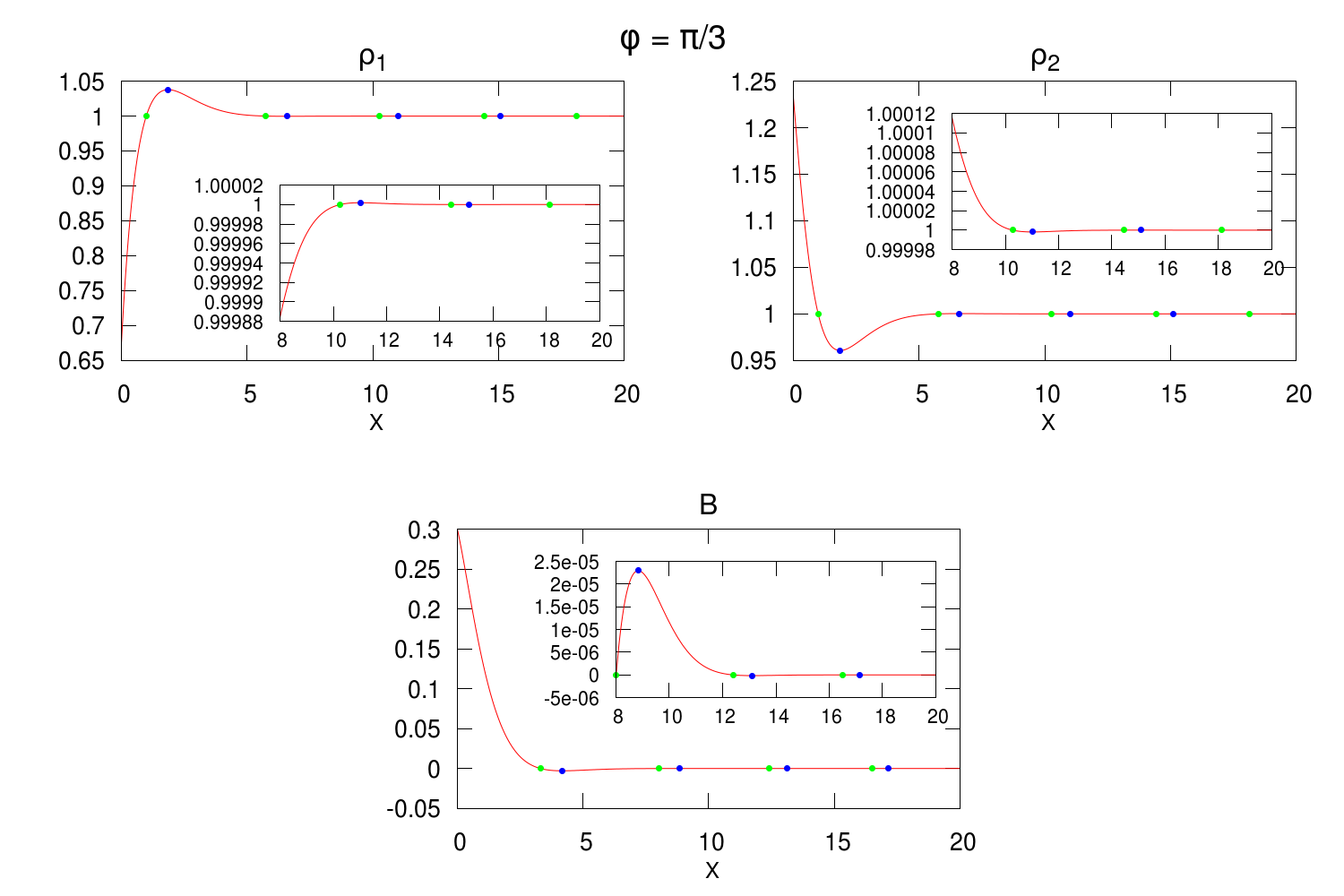}}}
 \caption{\label{Fig:Meissner1xi0} 
The Meissner state at a superconductor-insulator interface in the model \eqref{Eq:FpBS} with $V_0 = 3$, $\nu = -0.95$, $\chi=0$ and external field 
$H=0.3$ for two different boundary orientations: $\varphi=0$ (top set of plots) and $\varphi=\pi/3$ (bottom set of plots).
The boundary is at $X=0$, the plotted fields are the condensate magnitudes $\rho_1$ and $\rho_2$ and the magnetic field strength $B$. The green dots mark points where the fields cross their ground state values and the blue dots mark local extrema.
}
\end{figure*}

To confirm that the long-range decay behaviour holds for various boundary conditions,
we include here a plot in figure \ref{Fig:Meissner1xi0} for the parameters used in our results section but with the boundary term removed, $\chi = 0$.

\section{D: rescaling of fields}
We have made use of the form of the potential argued for in \cite{bouhon2010influence}, however we have made a few rescalings to rewrite the proposed model in a simpler fashion. The proposed model is,
\begin{eqnarray}
\nonumber
E^b &=& \int_{\mathbb{R}^2} \left\{ a_p (\left|\eta_x\right|^2 + \left|\eta_y\right|^2) 
+ b_1(\left|\eta_x\right|^2 + \left|\eta_y\right|^2)^2 
\right.\\ &&\left.
\nonumber
+ \frac{b_2}{2}(\overline{\eta}_x^2\eta_y^2 + \overline{\eta}_y^2 \eta_x^2 )
+b_3 \left|\eta_x\right|^2 \left|\eta_y\right|^2
\right.\\ &&\left.
\nonumber
+K_1(\left|D_1 \eta_x\right|^2 + \left|D_2 \eta_y\right|^2)
+ K_2( \left|D_1 \eta_y\right|^2 + \left|D_2 \eta_x\right|^2 
\right.\\ &&\left.
\nonumber
+ \overline{D_1 \eta_x} D_2 \eta_y + \overline{D_2 \eta_y} D_1 \eta_x + \overline{D_1 \eta_y} D_2 \eta_x + \overline{D_2 \eta_x} D_1 \eta_y)
\right.\\ &&\left.
+\frac{B^2}{8\pi}\right\}d^2 x^b.
\end{eqnarray}
where $D_i = \partial_i - i \gamma A^b_i$ and some of the parameters are coupled such that,
\begin{eqnarray}
K_1 = \frac{K}{4}(3+\nu) \quad \quad K_2 = \frac{K}{4}(1-\nu)\nonumber\\
b_1 = \frac{b}{8}(3+\nu) \quad \quad b_2 = \frac{b}{4}(1-\nu)\nonumber\\
b_3 = -\frac{b}{4}(1+3\nu). \quad \quad
\end{eqnarray}
We will write our condensate fields as,
\begin{equation}
\psi_1 = \eta_x / \lambda, \quad \psi_2 = \eta_2 / \lambda, \quad \lambda := \sqrt{-a_p / b},
\end{equation}
and rescale our gauge field,
\begin{equation}
A_i = A_i^b / \lambda_A, \quad \lambda_A := \lambda\sqrt{4\pi K}.
\end{equation}
Finally we can use a spatial rescaling,
\begin{equation}
x_i = x_i^b / \lambda_x, \quad \lambda_x := 1/\gamma \lambda_A,
\end{equation}
and then rescale the total energy to be,
\begin{equation}
E = E^b / \lambda_E, \quad \lambda_E := K \lambda^2 / 2.
\end{equation}
This finally gives the form of the energy given in equation \ref{Eq:F} with potential,
\begin{eqnarray}
&F_p& = V_0\left\{1 -(\rho_1^2+\rho_2^2)+ \frac{1}{8}\left(3+\nu\right)\left(\rho_1^2 + \rho_2^2\right)^2 \right. \nonumber \\  &-& \left. \frac{1}{4}\left(1+3\nu\right)\rho_1^2\rho_2^2  + \frac{1}{4}\left(1 - \nu\right)\rho_1^2\rho_2^2\cos{2\theta_{12}}\right\},
\end{eqnarray}
and anisotropy tensors,
\begin{gather*}
Q^{11} = \left(\begin{array}{cc} 3 + \nu & 0 \\ 0& 1-\nu\end{array}\right), \quad Q^{22} = \left(\begin{array}{cc} 1 - \nu & 0 \\ 0& 3 + \nu\end{array}\right), \\ \quad Q^{12} = \left(\begin{array}{cc} 0 & 1-\nu \\ 1-\nu & 0\end{array}\right).
\label{Eq:aniBS}
\end{gather*}
Where we have collected multiple parameters together,
\begin{equation}
V_0 = \frac{b}{2\pi \gamma^2 K^2}.
\end{equation}
Note that without loss of generality we have reduced the number of parameters to two $(V_0, \nu)$. This leads to the vacua and hence asymptotic values being $\theta_{12} = \pm \pi/2$ as required for BTRS and $\rho_1 = \rho_2 = 1$ without loss of generality.
\subsection*{Acknowledgements}
We thank Mihail Silaev for collaboration and discussions.
The work of MS and TW is supported by the UK Engineering and Physical Sciences Research Council through grant EP/P024688/1. EB is supported by the Swedish Research
Council Grants   No.\ 642-2013-7837, 2016-06122, 2018-03659
 and G\"{o}ran Gustafsson  Foundation  for  Research  in  Natural  Sciences  and  Medicine.
 This work was performed in part at the Aspen Center for Physics, which is supported by National Science Foundation grant PHY-1607611.

\bibliography{references}{}
 \end{document}